\documentclass[11pt,a4paper]{article}
\usepackage{jheppub}

\usepackage[latin1]{inputenc} 
\usepackage[T1]{fontenc}
\usepackage[safe]{textcomp}
\usepackage{lmodern} 
\usepackage{epsfig}
\usepackage{float}
\usepackage{amstext}  
\usepackage{amsfonts}
\usepackage{amssymb} 
\usepackage{amsmath}
\usepackage{amsthm}
\usepackage{bm}       
\usepackage[bottom]{footmisc} 
\usepackage{array}                
\usepackage{xcolor} 
\usepackage{framed}
\usepackage{slashed}
\usepackage[absolute]{textpos}
\usepackage{axodraw4j}
\usepackage{dsfont}
\usepackage{multirow}

\usepackage{appendix}
\usepackage{listings}
\usepackage{enumerate}     
\usepackage[bottom]{footmisc} 
\usepackage{array}           
\usepackage{parskip}        
\usepackage{xcolor} 
\usepackage{color}
\usepackage{framed}
\usepackage{fancyhdr}
\usepackage{subfig}

\usepackage{tikz}
\usepackage[absolute]{textpos}
\usepackage{pstricks}
\usepackage{axodraw4j}
\usepackage{graphicx}

\newcommand{\p}{\partial}

\newcommand{\f}[2]{\frac{#1}{#2}}
\newcommand{\sss}[1]{\scriptscriptstyle{#1}}
\newcommand{\ssst}[1]{\scriptscriptstyle{\text{#1}}}
\newcommand{\vv}[2]{\left( \begin{array}{c} #1 \\ #2  \end{array} \right)}
\newcommand{\bea}{\begin{eqnarray}}
\newcommand{\eea}{\end{eqnarray}}
\newcommand{\be}{\begin{equation}}
\newcommand{\ee}{\end{equation}}
\newcommand{\ba}{\begin{align}}
\newcommand{\ea}{\end{align}}
\newcommand{\beas}{\begin{eqnarray*}}
\newcommand{\eeas}{\end{eqnarray*}}
\newcommand{\bes}{\begin{equation*}}
\newcommand{\ees}{\end{equation*}}
\newcommand{\bas}{\begin{align*}}
\newcommand{\eas}{\end{align*}}

\newcommand{\ssL}{{\mathcal L}} 
\newcommand{\eps}{{\varepsilon}}
 
\newcommand{\cf}{C_{\scriptscriptstyle{F}}} 
\newcommand{\ca}{C_{\scriptscriptstyle{A}}}
\newcommand{\tr}{T_{\scriptscriptstyle{F}}}
\newcommand{\dF}{N_{\scriptscriptstyle{c}}}
\newcommand{\dR}{N_{\scriptscriptstyle{c}}}
\newcommand{\Ng}{n_{\scriptscriptstyle{g}}}

\newcommand{\Nf}{n_{\scriptscriptstyle{f}}}
\newcommand{\gs}{g_{\scriptscriptstyle{s}}}
\newcommand{\yt}{y_{\scriptscriptstyle{t}}}

\newcommand{\gb}{g_1}
\newcommand{\gw}{g_2}
\newcommand{\gaf}{\gamma_{\scriptscriptstyle{5}}}
\newcommand{\als}{\alpha_{\scriptscriptstyle{s}}}
\newcommand{\as}{a_{\scriptscriptstyle{s}}}
\newcommand{\at}{a_{\scriptscriptstyle{t}}}
\newcommand{\allam}{a_{\scriptscriptstyle{\lambda}}}
\newcommand{\lb}{\left(}
\newcommand{\rb}{\right)}

\newcommand{\msbar}{$\overline{\text{MS}}$}

\newcommand{\dFFfNg}{\frac{d_{\scriptscriptstyle{F}}^{abcd}d_{\scriptscriptstyle{F}}^{abcd}}{\Ng}}
\newcommand{\dFAfNg}{\frac{d_{\scriptscriptstyle{F}}^{abcd}d_{\scriptscriptstyle{A}}^{abcd}}{\Ng}}
\newcommand{\dAAfNg}{\frac{d_{\scriptscriptstyle{A}}^{abcd}d_{\scriptscriptstyle{A}}^{abcd}}{\Ng}}

\definecolor{bluemar}{rgb}{0,0,.5}
\definecolor{redmar}{rgb}{.8,0,0}
\definecolor{greenmar}{rgb}{0,.5,0}

\newcommand{\bk}{\color{black}}

\allowdisplaybreaks
\newcommand{\ice}[1]{\relax}

\title{Top-Yukawa effects on the $\beta$-function of the strong coupling in the SM at four-loop level}
\author[a]{M.~F.~Zoller}
\affiliation[a]{Institut f\"ur Theoretische Teilchenphysik, Karlsruhe
  Institute of Technology (KIT), \mbox{D-76128 Karlsruhe, Germany}}
\emailAdd{max.zoller@kit.edu}

\abstract{We present analytical results for the QCD $\beta$-function extended to the gaugeless limit
of the unbroken phase of the Standard Model at four-loop level. Apart from the strong coupling itself
we include the top-Yukawa contribution and the Higgs self-coupling. We observe a numerically small non-naive
$\gaf$ contribution at order $\yt^4\gs^4$, a feature not encountered in lower loop orders. We discuss the treatment of $\gaf$
which is more involved than in previous calculations at three-loop level.
}

\arxivnumber{}
\subheader{TTP15-031}

\begin{document}
\maketitle

\section{Introduction}

An important feature of the perturbative treatment of any quantum field theory is the evolution of couplings, fields and masses with
the renormalization scale $\mu$, which is usually set to a characteristic energy scale of the physical process under consideration.
This evolution is described by the Renormalization Group (RG) functions, i.~e.~$\beta$-functions for the couplings and anomalous dimensions for fields and masses.

The $\beta$-function for any coupling $X$ is defined as
\be
\beta_{\sss{X}}(X,X_1,X_2,\ldots)=\mu^2\f{d X}{d \mu^2}    =\sum \limits_{n=1}^{\infty} \f{1}{(16\pi^2)^{n}}\,\beta_{\sss{X}}^{(n)}{}.
\ee
It is a power series in all couplings \mbox{$X,X_1,X_2,\ldots$} of the theory
and independent of all gauge parameters $\xi$.

Recently the RG functions of the Standard Model (SM) were computed at three-loop accuracy. In the \msbar scheme $\beta$-functions do not depend on masses \cite{Collins:1974da}, hence
they can be computed in the unbroken phase of the SM.
For the gauge couplings $\gs,\gw$ and $\gb$ of the $\text{SU}_{\sss{C}}(3)$, $\text{SU}_{\sss{L}}(2)$ and
$\text{U}_{\sss{Y}}(1)$ subgroups of the SM the results were first published in \cite{PhysRevLett.108.151602,Mihaila:2012pz} and independently confirmed in \cite{Bednyakov:2012rb}.
For the top-Yukawa coupling $\yt$, which is the numerically largest Yukawa coupling by far, and the parameters of the Higgs potential 
$\lambda$ and $m^2$ the $\beta$-functions were first computed in the gaugeless limit, i.e. $\gw, \gb \to 0$, along with the anomalous dimensions of the fields involved \cite{Chetyrkin:2012rz}. 
Later $\beta_{\lambda}$ and $\beta_{m^2}$ were extended to the full SM \cite{Chetyrkin:2013wya}, confirmed by \cite{Bednyakov:2013eba,Bednyakov:2013cpa}, as well as
$\beta_{\yt}$ \cite{Bednyakov:2012en}, where the $\beta$-functions for the smaller Yukawa couplings were also added. 
The one- and two-loop $\beta$-functions for the gauge couplings
\cite{PhysRevLett.30.1343,PhysRevLett.30.1346,Jones1974531,Tarasov:1976ef,PhysRevLett.33.244,
Egorian:1978zx,PhysRevD.25.581,Fischler:1981is,Jack1985472,
Machacek198383,2loopbetayukawa,Ford:1992pn}, Yukawa couplings
\cite{Fischler1982385,Jack1985472,Machacek1984221,2loopbetayukawa} and Higgs potential parameters
\cite{Machacek198570,Jack1985472,2loopbetayukawa,Ford:1992pn}
have been known for a long time as well as partial three-loop results
\cite{Curtright:1979mg,Jones:1980fx,Tarasov:1980au,Tarasov1980429,3loopbetaqcd,Steinhauser:1998cm,Pickering:2001aq}. 
At four-loop level only the QCD $\beta$-function, i.e. $\beta_{\gs}(\gs)$ or equivalently 
$\beta_{\als}(\als)=2 \f{\als}{\gs}\beta_{\gs}$ with $\als=\f{\gs^2}{4\pi}$ is known \cite{4loopbetaqcd,Czakon:2004bu}.

Especially the evolution of the quartic Higgs self-coupling has received a lot of interest because of its close connection to the question of vacuum
stability in the Standard Model. It has been shown that the stability
of the SM vacuum up to some large energy scale $\Lambda \sim M_{\sss{Planck}}$ is approximately
equivalent to the requirement that the running coupling
$\lambda(\mu)>0$ for $\mu \leq \Lambda$ \cite{Cabibbo:1979ay,Ford:1992mv,Altarelli1994141}. 
The function $\beta_{\lambda}$ describing this evolution depends on all SM couplings an especially the large couplings $\yt$ and $\gs$ have a strong
influence. As the evolution of all couplings is interdependent a precision calculation for the evolution of all - at least of the five largest ($\gs$, $\yt$, $\gw$, $\gb$ and $\lambda$) -
is well motivated. Many analyses of this question have been performed
\cite{Bezrukov:2009db,Holthausen:2011aa,EliasMiro:2011aa,Xing:2011aa,Bezrukov:2012sa,Degrassi:2012ry,Chetyrkin:2012rz,Zoller:2012cv,Masina:2012tz,Zoller:2014cka,Zoller:2014xoa,Zoller:2013mra,Buttazzo:2013uya,Bednyakov:2015sca}
during the last years. 

In this paper we extend the QCD $\beta$-function to the gaugeless limit of the SM, i.~e.~we include the dependence on the top-Yukawa coupling $\yt$ and
the quartic Higgs self-coupling $\lambda$. This can be seen as a first step to all three gauge coupling $\beta$-functions in the full SM. 
To start with the gaugeless limit seems reasonable, first because at the energy-scales of our experiments $\yt$ is the second largest coupling in the SM after $\gs$,
followed by $\gw$, $\gb$ and $\lambda$. In order to renormalize fermion loops with four scalar legs we should also add counterterm $\propto \Phi^4$ to the Lagrangian of our simplified model.
This is exactly a contribution to the renormalization of $\lambda$ which makes it natural to include $\lambda$ as well. 

Secondly, the gaugeless limit of the SM provides an excellent opportunity to study the proper treatment of $\gaf$, which is introduced in the Yukawa-part of the Lagrangian.
This matrix is not well-defined in $D=4-2\eps$ dimensions and hence constitutes a non-trivial challenge.

The paper is structured as follows: In the following section the technical details, especially the treatment of $\gaf$, as well as
the automation of the calculation are discussed. Then the results are given and the relevance of the four-loop terms numerically determined at
the scale of the top quark mass. 

{\bf Note:} During the finishing process of this paper a similar calculation was published by another group  \cite{Bednyakov:2015ooa}.
Their calculation was not performed with massive tadpole integrals but rather with massless propagator-like integrals and in the Background field gauge. 
Both results achieved with different methods agree if the same prescription for the treatment of $\gaf$ is used (see section \ref{gafchap}).

\section{Details of the calculation}

\subsection{Gaugeless limit of the SM}

The Lagrangian of the SM in the unbroken phase can be decomposed into
\be
\ssL_{\ssst{SM}}=\ssL_{\ssst{SU(3)$\times$SU(2)$\times$U(1)}}+\ssL_{\ssst{Yukawa}}+\ssL_{\sss{\Phi}}{}, \label{LSM}
\ee
where $\ssL_{\ssst{SU(3)$\times$SU(2)$\times$U(1)}}$ contains the kinetic terms of the fermions and gauge bosons,
their interactions and the necessary gauge fixing and ghost terms. The Yukawa part $\ssL_{\ssst{Yukawa}}$ describes
the coupling of the fermions to a scalar SU(2) doublet $\Phi=\vv{\Phi_1}{\Phi_2}$ which results in fermion masses and the coupling 
of fermions to the Higgs boson after Spontaneous Symmetry Breaking as well as the mixing of the quark generations. 
The scalar part $\ssL_{\sss{\Phi}}$
contains the kinetic term for the scalar field $\Phi$, its potential and its coupling to the electroweak gauge bosons
through the covariant derivative. 
In the gaugeless limit we neglect two smaller gauge couplings $\gw$ and $\gb$ (electroweak sector).
We also approximate the small Yukawa couplings, i.~e.~all but the top-Yukawa coupling $\yt$, by zero and arrive at a simplified model
which includes QCD and top-Yukawa effects as well as the scalar potential:
\be
\ssL=\ssL_{\sss{QCD}}+\ssL_{\sss{\yt}}+\ssL_{\sss{\Phi}} \label{Lgaugeless}
\ee
with
\bea
\ssL_{\sss{QCD}}&=&-\f{1}{4}G^a_{\mu \nu} G^{a\,\mu \nu}-\f{1}{2 (1-\xi)}\lb\p_\mu A^{a\,\mu}\rb^2 
+\p_\mu \bar{c}^a \p^{\mu}c^a+\gs f^{abc}\,\p_\mu \bar{c}^a A^{b\,\mu} c^c \nonumber \\
&+&\sum\limits_q\left\{\f{i}{2}\bar{q}\overleftrightarrow{\slashed{\p}}q+ \gs \bar{q}\slashed{A}^a T^a q\right\}{},
\label{LQCD} \\
\ssL_{\sss{\yt}}
&=&-\yt \left\{ \lb \bar{t} P_{\sss{R}} t\rb \Phi^{*}_2+\lb\bar{t} P_{\sss{L}} t\rb \Phi_2
-\lb\bar{b} P_{\sss{R}} t\rb \Phi^{*}_1-\lb \bar{t} P_{\sss{L}} b\rb \Phi_1
\right\} {},
\label{LYuk} \\
\ssL_{\sss{\Phi}}&=&\p_\mu \Phi^\dagger \p^\mu \Phi-m^2\Phi^\dagger\Phi-\lambda \lb\Phi^\dagger\Phi\rb^2{}. 
\label{LPhi} 
\eea  
Here $q$ runs over all quark flavours, the gluon field strength tensor is given by
\be
G^a_{\mu \nu}=\p_\mu A^a_\nu - \p_\nu A^a_\mu + \gs f^{abc}A^b_\mu A^c_\nu
\ee
and $f^{abc}$ are the structure constants of the colour gauge group with the generators $T^a$ which satisfy
\be \left[ T^a,T^b \right]=if^{abc}T^c.\ee

The Yukawa sector mixes left-handed (L) and right-handed (R) Weyl spinors which can be projected out from Dirac spinors
used in our Feynman rules by the application of the projectors
\be 
P_{\sss{L}}=\f{1}{2}\lb 1-\gamma_5\rb \qquad P_{\sss{R}}=\f{1}{2}\lb 1+\gamma_5\rb
{}.
\ee
The left- and right-handed parts of the quark fields and vertices participating in the Yukawa interaction are renormalized
differently.

The Lagrangian \eqref{Lgaugeless} is renormalized with the counterterms
\bea
\delta\!\ssL_{\sss{QCD}}&=&-\f{1}{4}\delta\! Z^{(2g)}_3 \lb \p_\mu A^a_\nu - \p_\nu A^a_\mu \rb^2
-\f{1}{2}\delta\! Z^{(3g)}_1 \gs f^{abc}\lb \p_\mu A^a_\nu - \p_\nu A^a_\mu \rb A^b_\mu A^c_\nu \nonumber\\
&-&\f{1}{4}\delta\! Z^{(4g)}_1 \gs^2 \lb f^{abc} A^b_\mu A^c_\nu \rb^2
+\delta\! Z^{(2c)}_3 \p_\mu \bar{c}^a \p^{\mu}c^a+\delta\! Z^{(ccg)}_1 \gs f^{abc}\,\p_\mu \bar{c}^a A^{b\,\mu} c^c \label{LQCDc}\\
&+&\sum\limits_q\left\{
\f{i}{2}\bar{q}\overleftrightarrow{\slashed{\p}}\left[\delta\! Z^{(2q)}_{2,L}P_{\sss{L}}+\delta\! Z^{(2q)}_{2,R}P_{\sss{R}}\right]q
+\gs \bar{q}\slashed{A}^a T^a\left[\delta\! Z^{(qqg)}_{1,L}P_{\sss{L}} +\delta\! Z^{(qqg)}_{1,R}P_{\sss{R}}\right]q
\right\}{}, \nonumber \\
\delta\!\ssL_{\sss{Yukawa}}&=&-\delta\! Z^{(tb\Phi)}_1 
\yt\left\{ \lb \bar{t} P_{\sss{R}} t\rb \Phi^{*}_2+\lb\bar{t} P_{\sss{L}} t\rb \Phi_2
-\lb\bar{b} P_{\sss{R}} t\rb \Phi^{*}_1-\lb \bar{t} P_{\sss{L}} b\rb \Phi_1
\right\} {},
\label{LYukc} \\
\delta\!\ssL_{\sss{\Phi}}&=&\delta\! Z_2^{(2\Phi)}\p_\mu \Phi^\dagger \p^\mu \Phi-m^2\, \delta\! Z_{\Phi^2}\Phi^\dagger\Phi
+\delta\! Z_1^{(4\Phi)}\lb\Phi^\dagger\Phi\rb^2{}.
\label{LPhic} 
\eea
All these renormalization constants were computed at three-loop level in the course of the calculations in \cite{Chetyrkin:2012rz}.
The simplest way to derive the renormalization constant for the strong gauge coupling $\gs$ is via
\be 
Z_{\gs}=\f{Z^{(ccg)}_{1}}{Z^{(2c)}_{3}\sqrt{Z^{(2g)}_{3}}} \label{Zgscomputation}
\ee
where we use the renormalization constants $Z=1+\delta Z$ in the \msbar-scheme. 
All divergent integrals are regularized in $D=4-2 \eps$ space time dimensions.

\subsection{Automation and calculation with massive tadpoles}

The calculation begins with the generation of all necessary 1PI Feynman diagrams with two external ghost or gluon legs
for $Z^{(2c)}_3$ or $Z^{(2g)}_3$ and with two external ghost and one external gluon leg for $Z^{(ccg)}_1$. This was done
with the program QGRAF \cite{QGRAF}. 

The C++ programs Q2E and EXP \cite{Seidensticker:1999bb,Harlander:1997zb} are then used to identify the topology of the diagram. Later we will Taylor expand in the external
momenta and use projectors on the integrals in order to make them scalar. For example the ghost-gluon vertex corrections are
proportional to the outgoing ghost momentum $q^\mu$, where $\mu$ is the Lorentz index of the gluon leg. Hence we expand
to first order in $q$, use the projector $\f{q^\mu}{q^2}$ on the integral and set $q \to 0$ after that.
This is allowed as \msbar{} renormalization constants do not depend on external momenta. After having set all external momenta to zero
we are left with tadpole integrals. 
The fermion traces, the expansion in the external momenta and the insertion of counterterms in one-loop, 
two-loop and three-loop diagrams was performed using FORM \cite{Vermaseren:2000nd,Tentyukov:2007mu}. 
The colour factors were computed with the \mbox{FORM} package COLOR \cite{COLOR}. 
The tadpole integrals up to three-loop order were computed with
the \mbox{FORM}-based package \mbox{MATAD}\cite{MATAD}.

\begin{figure}[!ht]
\begin{center}
  \begin{tabular}{ll}
    \begin{picture}(200,120) (0,0)
    \SetWidth{0.5}
    \SetColor{Black}
    \ArrowLine(0,0)(0,100) 
    \ArrowLine(0,100)(150,100)
    \ArrowLine(150,100)(150,0) 
    \ArrowLine(150,0)(0,0)
    \ArrowLine(50,0)(50,50)
    \ArrowLine(100,0)(100,50)
    \ArrowLine(50,50)(50,100)
    \ArrowLine(100,50)(100,100)
    \ArrowLine(50,50)(100,50)
    \Vertex(50,0){2}
    \Vertex(50,50){2}
    \Vertex(50,100){2}
    \Vertex(100,0){2}
    \Vertex(100,50){2}
    \Vertex(100,100){2}
    \Text(-5,50)[rc]{\Large{{$p_1$}}}  
    \Text(75,108)[cc]{\Large{{$p_2$}}}
    \Text(155,50)[lc]{\Large{{$p_3$}}}
    \Text(75,8)[cc]{\Large{{$p_4$}}}
    \Text(45,25)[rc]{\Large{{$p_5$}}}
    \Text(45,75)[rc]{\Large{{$p_6$}}}
    \Text(105,75)[lc]{\Large{{$p_7$}}}
    \Text(105,25)[lc]{\Large{{$p_8$}}}
    \Text(75,58)[cc]{\Large{{$p_9$}}}
  \end{picture} &
      \begin{picture}(200,120) (0,0)
    \SetWidth{0.5}
    \SetColor{Black}
    \ArrowLine(0,0)(0,100) 
    \ArrowLine(0,100)(150,100)
    \ArrowLine(150,100)(150,0) 
    \ArrowLine(150,0)(0,0)
    \ArrowLine(50,0)(50,50)
    \ArrowLine(100,0)(100,50)
    \Line(50,50)(74,74)
    \Line(100,50)(75,75)
    \ArrowLine(76,76)(100,100)
    \ArrowLine(75,75)(50,100)    
    \ArrowLine(50,50)(100,50)
    \Vertex(50,0){2}
    \Vertex(50,50){2}
    \Vertex(50,100){2}
    \Vertex(100,0){2}
    \Vertex(100,50){2}
    \Vertex(100,100){2}
    \Text(-5,50)[rc]{\Large{{$p_1$}}}  
    \Text(75,108)[cc]{\Large{{$p_2$}}}
    \Text(155,50)[lc]{\Large{{$p_3$}}}
    \Text(75,8)[cc]{\Large{{$p_4$}}}
    \Text(45,25)[rc]{\Large{{$p_5$}}}
    \Text(60,80)[rc]{\Large{{$p_6$}}}
    \Text(90,80)[lc]{\Large{{$p_7$}}}
    \Text(105,25)[lc]{\Large{{$p_8$}}}
    \Text(75,58)[cc]{\Large{{$p_{10}$}}}
  \end{picture} \\ planar topology: tad4lp & non-planar topology: tad4lnp
  \end{tabular} \end{center}
  \caption{Four-loop tadpole topologies: $p_1$, $p_2$, $p_3$, $p_4$ are independent loop momenta, the others are linear combinations
\mbox{$p_5=p_4-p_1$},
\mbox{$p_6=p_2-p_1$},
\mbox{$p_7=p_3-p_2$},
\mbox{$p_8=p_3-p_4$},
\mbox{$p_9=p_4-p_2$} and
\mbox{$p_{10}=p_4+p_2-p_1-p_3$}.}
  \label{Topotad4l}
\end{figure}
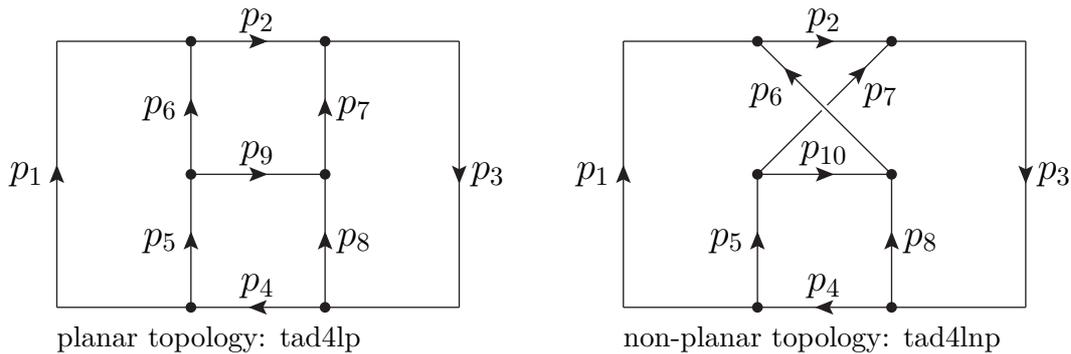

At four-loop level there are two independent tadpole topologies, see Fig.~\ref{Topotad4l}.
\footnote{All Feynman diagrams in this paper have been drawn with the 
Latex package Axodraw \cite{Vermaseren:1994je}.}
All scalar products $p_i\cdot p_j$ ($i,j=1,\ldots,10$) can be written as linear combinations of the $p_i^2$
which can be expressed in terms of the scalar propagators $D_i=\f{1}{i}\f{1}{M^2-p_i^2}$ and the auxiliary
Mass $M^2$ (see below). Hence all four-loop integrals can be written in terms of functions
\be \text{TAD4l}(n_1,\ldots,n_{10}):=\int\!\mathrm{d}^D\!p_1\int\!\mathrm{d}^D\!p_2\int\!\mathrm{d}^D\!p_3\int\!\mathrm{d}^D\!p_4
\prod\limits_{i=1}^{10} D_i^{n_i}{}. \label{unredInt}\ee

The integrals \eqref{unredInt} can be reduced to Master Intgrals (MI) using 
FIRE \cite{Smirnov:2008iw}. For the huge number of integrals in such a calculation the C++ version of
FIRE 5 \cite{Smirnov:2014hma} is necessary. All MI needed for this computation can be found in \cite{Czakon:2004bu}.
The program FIESTA 3 \cite{Smirnov:2013eza} was used to numerically cross check these MI and some unreduced
integrals as a check for our setup.

In order to compute the divergent part of the needed self-energies and vertex corrections we use
the same method as in our previous calculations \cite{Chetyrkin:2012rz,Chetyrkin:2013wya}.
This method was suggested in \cite{Misiak:1994zw} and further developed in \cite{beta_den_comp}. 
A step-by-step explanation of this method can be found in \cite{Zoller:2014xoa}.
An auxiliary mass parameter $M^2$ is introduced in every propagator denominator.
A naive Taylor expansion in the external momenta is performed before applying the projector to scalar integrals. 
After that all external
momenta are set to zero which leaves us with scalar tadpole integrals. Subdivergences $\propto M^2$ are canceled by counterterms
\be \begin{split}
\f{M^2}{2}\delta\!Z_{\sss{M^2}}^{(2g)}\,A_\mu^a A^{a\,\mu}\;
\text{and} \;
\f{M^2}{2}\delta\!Z_{\sss{M^2}}^{(2\Phi)}\, \Phi^\dagger\Phi{}.
\end{split} \ee
which are computed order by order in perturbation theory and inserted in lower loop diagrams.
Note that this method is only valid for computing UV divergent parts of Feynman diagrams, and hence Z-factors, not finite amplitudes.

\subsection{Treatment of $\gaf$} \label{gafchap}
The most important issue of this calculation is the proper treatment of $\gaf$ in dimensional regularization. In $D=4$ dimensions it can be defined as 
\be \gamma_5=i\gamma^0\gamma^1\gamma^2\gamma^3=\f{i}{4!}\eps_{\mu\nu\rho\sigma} \gamma^\mu\gamma^\nu\gamma^\rho\gamma^\sigma \text{ with }
 \eps_{0123}=1=-\eps^{0123} \label{gamma5} 
{}.
\ee
In most diagrams a naive treatment of $\gaf$ is sufficient, i.~e.~we use $\left\{\gaf,\gamma^\mu\right\}=0$ and $\gaf^2=1$,
valid in $D=4$ dimensions, until only
one or no $\gaf$ matrix remain on each fermion line, then discard diagrams with at least one $\gaf$.
This is valid for fermion lines with less than four Lorentz indices and momenta flowing into the fermion line.
\begin{figure}
 \begin{tabular}{cc}
 \scalebox{0.8}{
  \begin{picture}(230,45) (0,0) 
\ArrowArc(100,0)(50,180,360)
\ArrowArc(100,0)(50,0,180)
\Gluon(50,0)(0,0){6}{3}
\Gluon(150,0)(200,0){6}{3}
\Gluon(65,-35)(45,-75){6}{3}
\Gluon(135,-35)(165,-75){6}{3}
\DashArrowLine(45,75)(65,35){4}
\DashArrowLine(165,75)(135,35){4}
\SetColor{Red}
\Vertex(65,35){4}
\LongArrow(35,75)(55,35)
\LongArrow(175,75)(145,35)
 \Text(75,15)[cc]{\Large{\Red{$\gamma_5$}}}
\Text(0,-16)[cc]{{\Red{$\mu_1$}}}
\Text(200,-16)[cc]{{\Red{$\mu_2$}}}
\Text(45,-91)[cc]{{\Red{$\mu_3$}}}
\Text(165,-91)[cc]{{\Red{$\mu_4$}}}
\Text(25,85)[cc]{{\Red{$k_2$}}}
\Text(185,85)[cc]{{\Red{$k_1$}}}
\end{picture}}
& \scalebox{0.8}{
\begin{picture}(230,45) (0,0) 
 \ArrowLine(0,0)(200,0)
 \ArrowLine(180,0)(230,0)
 \ArrowLine(0,0)(50,0)
 \Gluon(50,0)(50,40){4}{4}
 \Gluon(80,0)(80,40){4}{4}
 \Gluon(120,0)(120,40){4}{4}
 \Gluon(180,0)(180,40){4}{4}
 \DashArrowLine(150,0)(150,40){4}
 \COval(115,50)(20,75)(0){Black}{Blue}
\Text(100,15)[cc]{{\Black{$\cdots$}}}
\SetColor{Red}
\Vertex(150,0){4}
\LongArrowArc(180,10)(30,220,320)
 \Text(180,-35)[cc]{\Large{\Red{$\gamma_5$}}}
\end{picture}}\vspace{20mm}\\
(a) & (b)
\end{tabular}
\caption{$\gaf$ on internal (a) and external (b) fermion lines}
\label{gammafiveintext}
\end{figure}
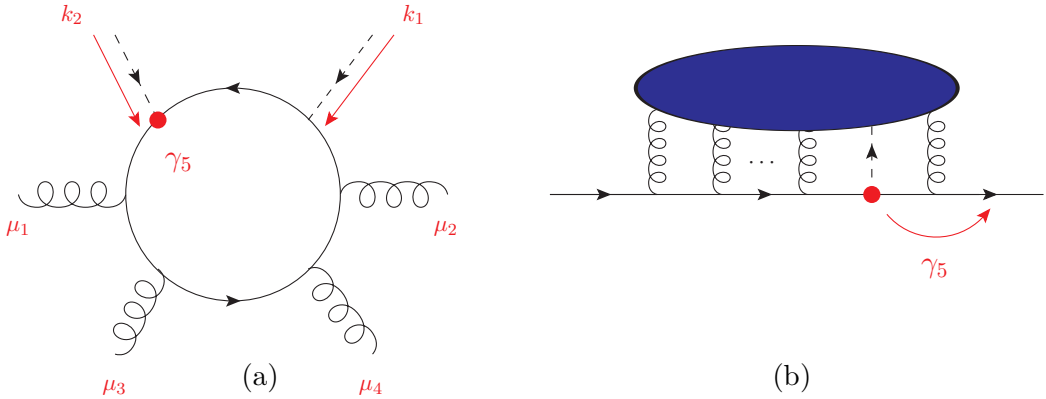
Fig.~\ref{gammafiveintext} shows the schematic cases of $\gaf$ appearing on internal and external fermion lines.
We start with the case of internal lines (see Fig.~\ref{gammafiveintext} (a)). In fact, for the calculation presented in this paper no external fermion lines appear.

As we set all momenta external to the whole Feynman diagram to zero for the computation of the UV divergent part
of the diagram external momenta to a fermion line ($k_1$, $k_2$,$\ldots$) are loop momenta from other loops.
Taking the trace over the closed fermion loop in $D=4$ dimensions yields a result with terms    
proportional to $\eps_{\mu_1\mu_2\mu_3\mu_4}$ and $\eps_{\mu_1 \mu_2 \alpha \beta}\, k_1^\alpha\, k_2^\beta$ and so on.
In order for the $\eps$-tensors not to vanish at least 4 free Lorentz structures are needed.
Else the diagram is set to zero.

If we have only one internal fermion line with one $\gaf$ on it and the final result is known to be scalar (not pseudoscalar), as are the counterterms we want to compute here,
we can discard these terms as well. The only possibility for a non-naive contribution to the final result can appear in the case
of two (or more) fermion lines. Here the two $\eps$-tensors can be contracted and expressed in terms of the metric tensor
\be \eps^{\mu_1\mu_2\mu_3\mu_4} \eps_{\nu_1\nu_2\nu_3\nu_4}
=-\sum\limits_{\pi}\,\text{sgn}(\pi) g^{\mu_{\pi(1)}}_{\,\,\,\,\nu_1} g^{\mu_{\pi(2)}}_{\,\,\,\,\nu_2} g^{\mu_{\pi(3)}}_{\,\,\,\,\nu_3} 
g^{\mu_{\pi(4)}}_{\,\,\,\,\nu_4}
{}, \label{epscontraction} \ee
where the sum is taken over all permutations $\pi$ of (1,2,3,4) and \be \text{sgn}(\pi)=\begin{cases}
                                                                                      +1 & \text{ for $\pi$ even} \\ -1 & \text{ for $\pi$ odd}
                                                                                     \end{cases}
{}. \ee
The lhs of \eqref{epscontraction} is composed of intrinsically four-dimensional objects whereas the rhs can be used in $D=4-2\eps$ dimensions,
introducing an uncertainty of $\mathcal{O}(\eps)$. However, if the integrals appearing in the calculation of the Feynman diagram in question have only
$\f{1}{\eps}$ poles the divergent part, which we are interested in here, is unaffected.

For completeness we want to make a short remark about external fermion lines, such as the one shown in Fig.~\ref{gammafiveintext} (b), as well.
Here we can anticommute the $\gaf$ to the end of the fermion line and hence outside of all loops. But if we use a projector on the external fermion line
in order to make the integral scalar and this involves taking a trace over the fermion line we have to treat it the same way as the
internal ones. In the case of the three-loop $\beta$-function for the Yukawa couplings a non-naive $\gaf$ effect from the contraction
of the $\eps$-tensors from an internal and an external fermion line was observed \cite{Chetyrkin:2012rz}.

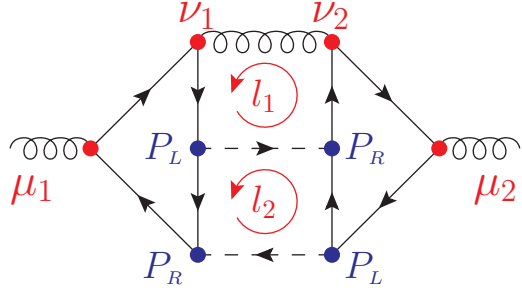
\begin{figure}[!ht]
  \begin{center}
    \begin{picture}(200,110) (0,-50)
    \SetWidth{0.5}
    \SetColor{Black}
    \Gluon(0,0)(30,0){4}{3}
    \ArrowLine(30,0)(70,40)
    \ArrowLine(70,-40)(30,0)
    \ArrowLine(70,40)(70,0)
    \ArrowLine(70,0)(70,-40)
        \Gluon(70,40)(120,40){4}{5}
        \DashArrowLine(70,0)(120,0){6}
        \DashArrowLine(120,-40)(70,-40){6}
    \ArrowLine(120,40)(160,0)
    \ArrowLine(160,0)(120,-40)
    \ArrowLine(120,0)(120,40)
    \ArrowLine(120,-40)(120,0)
    \Gluon(160,0)(190,0){4}{3}
\SetColor{Red}
        \Text(0,-10)[lt]{\LARGE{\Red{$\mu_1$}}}    
        \Text(190,-10)[rt]{\LARGE{\Red{$\mu_2$}}}
        \Vertex(30,0){3}
        \Vertex(160,0){3}
                \Text(70,45)[cb]{\LARGE{\Red{$\nu_1$}}}    
                \Text(120,45)[cb]{\LARGE{\Red{$\nu_2$}}}
                \Vertex(70,40){3}
                \Vertex(120,40){3}
\LongArrowArc(95,20)(12,210,160)
\Text(95,20)[cc]{\Large{\Red{$l_1$}}}
\LongArrowArc(95,-20)(12,210,160)
\Text(95,-20)[cc]{\Large{\Red{$l_2$}}}
\SetColor{Blue}
        \Vertex(70,0){3}
        \Vertex(120,0){3}
        \Vertex(70,-40){3}
        \Vertex(120,-40){3}
        \Text(65,0)[rc]{\Large{\Blue{$P_{\sss{L}}$}}}
        \Text(65,-45)[rc]{\Large{\Blue{$P_{\sss{R}}$}}}
        \Text(125,0)[lc]{\Large{\Blue{$P_{\sss{R}}$}}}
        \Text(125,-45)[lc]{\Large{\Blue{$P_{\sss{L}}$}}}
\SetColor{Black}
  \end{picture}
\end{center}
\caption{Diagram giving a non-naive $\gamma_5$-contribution to the gluon self-energy: each fermion line has two indices $\mu_i$ and $\nu_i$ (i=1,2) and two momenta
$l_1, l_2$, which can support a $\eps^{\mu_i \nu_i \alpha \beta}l_\alpha l_\beta$ term. The left- and right-handed projectors $P_{\sss{L,R}}$ introduce $\gaf$ into the diagram.} 
\label{diasggG5ep}
\end{figure}

In the calculations needed for the renormalization constants in \eqref{Zgscomputation} only one type of diagram features two fermion lines with four external
Lorentz indices or loop-momenta to them, namely in the gluon propagator, when each external leg is attached to a different fermion loop and the two fermion loops are
connected by a gluon and two $\Phi$-lines. A planar example is shown in Fig.~\ref{diasggG5ep}

There are 72 diagrams contributing to the non-naive part of the gluon propagator, which (like Fig.~\ref{diasggG5ep}) are all obtained by connecting two fermion loops
with an external gluon leg each by means of one gluon propagator and two scalar propagators in all possible ways.
Using $\{\gaf,\gamma^\mu\}=0$ we move all $\gaf$ matrices on each fermion line to the same reading point, for which we choose the external vertex. 
We checked that the same result is obtained if we choose to place $\gaf$ to the left or to the right of the external $\gamma^{\mu_{1,2}}$.
We can also use the Larin prescription \cite{Larin:1993tq}
\be \gamma^\mu\gaf=\f{i}{3!}\eps^{\mu\rho_1\rho_2\rho_3}\gamma_{\rho_1}\gamma_{\rho_2}\gamma_{\rho_3}{}, \ee
which combines the two possibilities, with the same result. It is only important that the reading point is the same for all 72 diagrams.
Due to $\gaf^2=1$ we are left with one or no $\gaf$ on each fermion line. If there is only one $\gaf$ on one fermion line the contribution is zero.
Terms with no $\gaf$ contribute to the naive part of the gluon propagator. The remaining contribution from one $\gaf$ on each fermion line is what we call the non-naive
contribution. The $\gaf$ prescription using the same external vertex in all diagrams was described in \cite{Korner:1991sx} as a practical and consistent
$\gaf$ scheme.

We checked explicitly that only $\f{1}{\eps}$ poles appear in the results for these diagrams. In fact, as an additional precaution we checked that
at $\mathcal{O}(\eps)$ completely antisymmetric and completely symmetric structures composed of the metric and the eight indices appearing in the $\eps$ tensors do not give contributions to
the divergent part. This was implemented as
\bea 
\eps^{\mu_1\mu_2\mu_3\mu_4} \eps_{\nu_1\nu_2\nu_3\nu_4} = 
&-&\sum\limits_{\pi}\,\text{sgn}(\pi) g^{\mu_{\pi(1)}}_{\,\,\,\,\nu_1} g^{\mu_{\pi(2)}}_{\,\,\,\,\nu_2} g^{\mu_{\pi(3)}}_{\,\,\,\,\nu_3} 
g^{\mu_{\pi(4)}}_{\,\,\,\,\nu_4}
\lb 1+\eps \cdot \text{C}^{\text{as}}\rb \nonumber\\
&+& \eps \cdot \text{C}^{\text{s}}\sum\limits_{\pi}\, g^{\mu_{\pi(1)}}_{\,\,\,\,\nu_1} g^{\mu_{\pi(2)}}_{\,\,\,\,\nu_2} g^{\mu_{\pi(3)}}_{\,\,\,\,\nu_3} g^{\mu_{\pi(4)}}_{\,\,\,\,\nu_4}{},
\eea
where the labels $\text{C}^{\text{as,s}}$ parametrize the uncertainty introduced through \eqref{epscontraction} being applied in $D=4-2\eps$. As they drop out in the divergent term of our final result
we are convinced that $\gaf$ can be treated in this way. 

However, in contrast to the Yukawa coupling $\beta$-functions at three-loop level, we
find here that the result is different if we do not choose the same reading point for $\gaf$ before taking the trace. 

For instance, if we leave
each $\gaf$ matrix at the point on the fermion line where it was introduced by the Feynman rules, i.~e.~we do not use $\left\{\gaf,\gamma^\mu\right\}=0$
at all in terms with one $\gaf$ on each fermion line, the result for these terms is a factor $3$ larger.\footnote{In the previous version of this paper a factor $6$ was given due to
a bug in this particular calculation. Thanks to the authors of \cite{Bednyakov:2015ooa} for pointing out the discrepancy with their calculation of the same quantity.}
This procedure is the opposite of moving all $\gaf$ to a
common reading point, but note that
we still use $\{\gaf,\gamma^\mu\}=0$ and $\gaf^2=1$ in terms with two $\gaf$ on one fermion line. 
This shows, however, that anticommuting $\gaf$ along the fermion
lines arbitrarily in each diagram spoils the result even though only $\f{1}{\eps}$ poles are visible in the final result. 
This becomes clear when we use $D=4-2 \tilde{\eps}$
when evaluating the fermion traces and $D=4-2 \eps$ in the intergral reduction and the master integrals. Then we see terms $\propto\f{\tilde{\eps}}{\eps^2}$
independent of the labels $\text{C}^{\text{as,s}}$. This means that the ambiguity is introduced by anticommuting the $\gaf$ to different points in different terms.
At present this issue is not fully understood. The approach described above using the external reading point seems intuitive. The result is also stable 
for choices of the reading point to the left or right of the external vertex. We check that the numerical impact of the non-naive terms is small. 
In fact, even a non-naive contribution of a factor $3$ larger would be numerically small compared to the naive contribution.

Naturally, we checked that this treatment of $\gaf$ respects the Ward identity manifest in the transversal structure of the gluon self-energy.

\section{Results \label{res:beta}}
In this section we give the results for the four-loop $\beta$-function of the strong coupling $\gs$ in the gaugeless limit of the SM.
For a gerneric SU($\dF$) gauge group the colour factors are expressed
through the quadratic Casimir operators $\cf$ and $\ca$ of the 
fundamental and the adjoint representation of the corresponding Lie algebra.
The dimension of the fundamental representation is called $\dF$. The adjoint representation has dimension $\Ng$ and the trace
$\tr$ defined by \mbox{$\tr \delta^{ab}=\textbf{Tr}\lb T^a T^b\rb$}  
with the group generators $T^a$ of the fundamental representation. In addition we need a few higher order invariants
constructed from the symmetric tensors
\bea
d_{\scriptscriptstyle{F}}^{abcd} &=& \f{1}{6} \text{Tr} 
\lb T^a T^b T^c T^d + T^a T^b T^d T^c + T^a T^c T^b T^d \right. \nonumber \\
&+& \left. T^a T^c T^d T^b + T^a T^d T^b T^c + T^a T^d T^c T^b \rb{}.
\eea
from the generators of the fundamental representation and analogously $d_{\scriptscriptstyle{A}}^{abcd}$ constructed from the generators of the adjoint representation.
The combinations needed and their SU($\dF$) values are
\bea
\dFFfNg &=& \frac{\dF^4-6\dF^2+18}{96\dF^2},\;\;\;\;
\dFAfNg = \frac{\dF(\dF^2+6)}{48},\\
\dAAfNg &=& \frac{\dF^2(\dF^2+36)}{24}.\nonumber
\eea
Furthermore for SU($\dF$) we have
\be \tr = \f{1}{2},\;\;\;\; \cf = \f{\dF^2-1}{2\dF},\;\;\;\; \ca = \dF,\;\;\;\; \Ng = \dF^2-1{}.\ee
The number of active fermion flavours is denoted by $\Nf$ (=6 in the SM).
\be
\begin{split}
\f{\beta_{\sss{\gs}}^{(4)}}{\gs}= &
         \gs^8   \left(
           \f{40}{9} \dAAfNg
          - \f{150653}{972} \ca^4
          - \f{256}{9} \Nf \dFAfNg
          - 23 \Nf \tr \cf^3     \right. \\ & \left.
          + \f{2102}{27} \Nf \ca \tr \cf^2
          - \f{7073}{486} \Nf \ca^2 \tr \cf
          + \f{39143}{162} \Nf \ca^3 \tr
          + \f{352}{9} \Nf^2 \dFFfNg     \right. \\ & \left.
          - \f{676}{27} \Nf^2 \tr^2 \cf^2
          - \f{8576}{243} \Nf^2 \ca \tr^2 \cf
          - \f{3965}{81} \Nf^2 \ca^2 \tr^2
          - \f{616}{243} \Nf^3 \tr^3 \cf     \right. \\ & \left.
          - \f{212}{243} \Nf^3 \ca \tr^3
          - \f{352}{3} \zeta_{3} \dAAfNg
          + \f{22}{9} \zeta_{3} \ca^4
          + \f{832}{3} \zeta_{3} \Nf \dFAfNg     \right. \\ & \left.
          - \f{176}{9} \zeta_{3} \Nf \ca \tr \cf^2
          + \f{328}{9} \zeta_{3} \Nf \ca^2 \tr \cf
          - \f{68}{3} \zeta_{3} \Nf \ca^3 \tr
          - \f{256}{3} \zeta_{3} \Nf^2 \dFFfNg     \right. \\ & \left.
          + \f{352}{9} \zeta_{3} \Nf^2 \tr^2 \cf^2
          - \f{224}{9} \zeta_{3} \Nf^2 \ca \tr^2 \cf
          - \f{112}{9} \zeta_{3} \Nf^2 \ca^2 \tr^2
          \right) \\ &
           + \gs^6 \yt^2   \left(
          - 3 \tr \cf^2
          - \f{523}{18} \ca \tr \cf
          - \f{985}{9} \ca^2 \tr
          + \f{322}{9} \Nf \tr^2 \cf     \right. \\ & \left.
          + \f{218}{9} \Nf \ca \tr^2
          + 72 \zeta_{3} \tr \cf^2
          + 36 \zeta_{3} \ca \tr \cf
          \right) \\ &
                 + \gs^4 \yt^4   \left(
          - 3 \tr \cf
          + \f{41}{2} \tr \cf \dR
          + 36 \ca \tr
          + 25 \ca \tr \dR \right. \\ & \left.
          - 24 \zeta_{3} \tr \cf \dR    
          +\fcolorbox{blue}{white}{$\tr^2\lb
           \f{80}{3} 
          - 32 \zeta_{3}  \rb$}\;
          \right) \\ &
                + \gs^2 \yt^6   \left(
          - \f{21}{4} \tr
          - 29 \tr \dR
          - \f{3}{2} \tr \dR^2
          - 6 \zeta_{3} \tr
          \right)  \\ &
       - 30 \gs^2 \yt^4 \lambda \tr  
        + 36 \tr \gs^2 \yt^2 \lambda^2 .
\end{split}
\label{4lbetags}
\ee
This is in agreement with \cite{Bednyakov:2015ooa} if the same $\gaf$ prescription is used. The term $\propto\gs^4 \yt^4 \tr^2$ is the only one affected
by non-naive $\gaf$ contributions as explained above. The naive and non-naive (i.~e.~stemming from the contraction of two $\eps$-tensors)
contributions are
\be 
\gs^4 \yt^4  \tr^2 \lb \f{80}{3} 
          - 32 \zeta_{3} \rb
=  \gs^4 \yt^4  \tr^2  \lb     
          \underbrace{ 24 }_{\text{ (naive)}}
          + \underbrace{ \f{8}{3} }_{\text{ (non-naive)}}
          - \underbrace{ 48 \zeta_{3}}_{\text{ (naive)}}
          + \underbrace{ 16 \zeta_{3}}_{\text{ (non-naive)}} \rb. \ee
The lower loop results are
\bea
\f{\beta_{\sss{\gs}}^{(3)}}{\gs} &=&
           \gs^6   \left(
          - \f{2857}{108} \ca^3
          - \Nf \tr \cf^2
          + \f{205}{18} \Nf \ca \tr \cf  \right. \nonumber \\ & &\left.
          + \f{1415}{54} \Nf \ca^2 \tr
          - \f{22}{9} \Nf^2 \tr^2 \cf
          - \f{79}{27} \Nf^2 \ca \tr^2
          \right)\\ & &
           -  \gs^4 \yt^2   \left(
           3 \tr \cf
          + 12 \ca \tr
          \right) 
                +  \gs^2 \yt^4   \left(
          + \f{9}{2} \tr
          + \f{7}{2} \tr \dR
          \right), \nonumber \\
\f{\beta_{\sss{\gs}}^{(2)}}{\gs} &=&  
        \gs^4   \left(
          - \f{17}{3} \ca^2
          + 2 \Nf \tr \cf
          + \f{10}{3} \Nf \ca \tr
          \right)       -2 \gs^2 \yt^2, \\
\f{\beta_{\sss{\gs}}^{(1)}}{\gs} &=& 
        \gs^2   \left(
          - \f{11}{6} \ca
          + \f{2}{3} \Nf \tr
          \right).
\eea
in agreement with \cite{Chetyrkin:2012rz}.
The pure QCD part of \eqref{4lbetags} agrees with \cite{4loopbetaqcd,Czakon:2004bu}.

For convenience we also give the $\beta$-function for $\als$. We absorb the loop factor $\f{1}{16\pi^2}$ into
\be \as=\f{\gs^2}{(4\pi)^2}=\f{\als}{4\pi},\quad \at=\f{\yt^2}{(4\pi)^2},\quad \allam=\f{\lambda}{(4\pi)^2}{} \ee and define
\be \beta_{\als}(\as,\at,\allam)=\sum \limits_{n=1}^{\infty} \beta_{\sss{\als}}^{(n)}(\as,\at,\allam).\ee
We find
\be
\begin{split}
\f{\beta_{\sss{\als}}^{(4)}}{\als}=& 
        \as^4   \left(
           \f{80}{9} \dAAfNg
          - \f{150653}{486} \ca^4
          - \f{512}{9} \Nf \dFAfNg
          - 46 \Nf \tr \cf^3      \right. \\ & \left.
          + \f{4204}{27} \Nf \ca \tr \cf^2 
          - \f{7073}{243} \Nf \ca^2 \tr \cf
          + \f{39143}{81} \Nf \ca^3 \tr  \right. \\ & \left.
          + \f{704}{9} \Nf^2 \dFFfNg     
          - \f{1352}{27} \Nf^2 \tr^2 \cf^2
          - \f{17152}{243} \Nf^2 \ca \tr^2 \cf \right. \\ & \left.
          - \f{7930}{81} \Nf^2 \ca^2 \tr^2
          - \f{1232}{243} \Nf^3 \tr^3 \cf     
          - \f{424}{243} \Nf^3 \ca \tr^3
          - \f{704}{3} \zeta_{3} \dAAfNg \right. \\ & \left.
          + \f{44}{9} \zeta_{3} \ca^4 
          + \f{1664}{3} \zeta_{3} \Nf \dFAfNg     
          - \f{352}{9} \zeta_{3} \Nf \ca \tr \cf^2\cf\right. \\ & \left.
          + \f{656}{9} \zeta_{3} \Nf \ca^2 \tr 
          - \f{136}{3} \zeta_{3} \Nf \ca^3 \tr 
          - \f{512}{3} \zeta_{3} \Nf^2 \dFFfNg   \cf\right. \\ & \left.  
          + \f{704}{9} \zeta_{3} \Nf^2 \tr^2 \cf^2
          - \f{448}{9} \zeta_{3} \Nf^2 \ca \tr^2 \cf
          - \f{224}{9} \zeta_{3} \Nf^2 \ca^2 \tr^2
          \right) \\ &
       + \at \as^3   \left(
          - 6 \tr \cf^2
          - \f{523}{9} \ca \tr \cf
          - \f{1970}{9} \ca^2 \tr
          + \f{644}{9} \Nf \tr^2 \cf \right. \\ & \left.
          + \f{436}{9} \Nf \ca \tr^2
          + 144 \zeta_{3} \tr \cf^2
          + 72 \zeta_{3} \ca \tr \cf
          \right) \\ &
       + \at^2 \as^2   \left(
          - 6 \tr \cf
          + 41 \tr \cf \dR
          + 72 \ca \tr
          + 50 \ca \tr \dR  \right. \\ & \left.
          - 48 \zeta_{3} \tr \cf \dR
          + \fcolorbox{blue}{white}{$\tr^2 \lb  \f{160}{3} 
          - 64 \zeta_{3}  \rb$}\;
          \right) \\ &
       + \at^3 \as   \left(
          - \f{21}{2} \tr
          - 58 \tr \dR
          - 3 \tr \dR^2
          - 12 \zeta_{3} \tr
          \right)\\ &
                 + \at^2 \as \allam   \left(
          - 60 \tr
          \right)        + \at \as \allam^2   \left(
          + 72 \tr
          \right),
\label{4lbetaas}
\end{split}
\ee
where
\be \at^2 \as^2 \tr^2 \lb + \f{160}{3} \tr^2 - 64 \zeta_{3} \tr^2 \rb
= \at^2 \as^2 \tr^2   \lb       \underbrace{ 48 }_{\text{(naive)}}
          + \underbrace{ \f{16}{3}  }_{     \text{(non-naive)}}
          - \underbrace{  96 \zeta_{3} }_{  \text{(naive)}}
          + \underbrace{  32 \zeta_{3}  }_{ \text{(non-naive)}} \rb. \ee
and
\bea
\f{\beta_{\sss{\als}}^{(3)}}{\als} &=& 
        \as^3   \left(
          - \f{2857}{54} \ca^3
          - 2 \Nf \tr \cf^2
          + \f{205}{9} \Nf \ca \tr \cf \right. \nonumber \\ & &\left.
          + \f{1415}{27} \Nf \ca^2 \tr
          - \f{44}{9} \Nf^2 \tr^2 \cf
          - \f{158}{27} \Nf^2 \ca \tr^2
          \right)\\ & &
                 + \at \as^2   \left(
          - 6 \tr \cf
          - 24 \ca \tr
          \right)
                 + \at^2 \as   \left(
          + 9 \tr
          + 7 \tr \dR
          \right), \nonumber \\
\f{\beta_{\sss{\als}}^{(2)}}{\als} &=& 
        \as^2   \left(
          - \f{34}{3} \ca^2
          + 4 \Nf \tr \cf
          + \f{20}{3} \Nf \ca \tr
          \right) 
                 + \at \as   \left(
          - 4 \tr
          \right), \\
\f{\beta_{\sss{\als}}^{(1)}}{\als} &=& 
        \as   \left(
          - \f{11}{3} \ca
          + \f{4}{3} \Nf \tr
          \right) {}.
\eea

Now we want to give a numerical evaluation of the $\beta$-functions at the scale of the top mass in order to get an idea
of the size of the new terms.
For \mbox{$M_t \approx 173.34\pm 0.76$ GeV} \cite{ATLAS:2014wva}, \mbox{$M_H \approx 125.09\pm 0.24$ GeV}\cite{Aad:2015zhl} 
and \mbox{$\als(M_Z)=0.1184\pm 0.0007$} \cite{Bethke:2012jm}
we get the couplings in the $\overline{\text{MS}}$-scheme at this scale using two-loop matching relations \cite{Buttazzo:2013uya}
\bea 
\gs(M_t)&=&1.1666 \pm 0.0035 \text{(exp)}, \nonumber \\
\yt(M_t)&=&0.9369 \pm 0.0046 \text{(exp)} \pm 0.0005 \text{(theo)}, \\
\lambda(M_t)&=&0.1259 \pm 0.0005 \text{(exp)} \pm 0.0003 \text{(theo)} \nonumber
\eea
where the experimental uncertainty (exp) stems from $M_t, M_H$ and $\als(M_Z)$ and the theoretical one (theo) from the matching of on-shell to \msbar{} parameters
(these are taken from \cite{Buttazzo:2013uya}). We find\footnote{The labels under the braces indicate from which part
of the $\beta$-function the contributions come.}
\bea
\f{\beta_{\sss{\gs}}^{(2)}}{\beta_{\sss{\gs}}^{(1)}(16\pi^2)}
&=& \underbrace{3.20\times 10^{-2}}_{\gs^4} 
 \underbrace{+1.59\times 10^{-3}}_{\gs^2\yt^2}, \\
\f{\beta_{\sss{\gs}}^{(3)}}{\beta_{\sss{\gs}}^{(1)}(16\pi^2)^2}
&=& \underbrace{ -3.45\times 10^{-4} }_{\gs^6} 
\underbrace{ +2.74\times 10^{-4} }_{\gs^4\yt^2} 
\underbrace{ -6.62\times 10^{-5} }_{\gs^2\yt^4} , \\
\f{\beta_{\sss{\gs}}^{(4)}}{\beta_{\sss{\gs}}^{(1)}(16\pi^2)^3}
&=&
\underbrace{2.26  \times 10^{-4}}_{\gs^8}
\underbrace{+2.47  \times 10^{-5}}_{\gs^6\yt^2}
\underbrace{-1.06  \times 10^{-5}}_{\gs^4 \yt^4 (\text{naive})}
\underbrace{-4.17 \times 10^{-7}}_{\gs^4\yt^4 (\text{non-naive})}\\ & &
\underbrace{ +2.77 \times 10^{-6}}_{\gs^2\yt^6} 
\underbrace{+ 1.06 \times 10^{-7}}_{\gs^2\yt^4\lambda}
\underbrace{-1.82  \times 10^{-8}}_{\gs^2\yt^2\lambda^2}
\eea
We see that the top-Yukawa contributions have a sizable impact on the four-loop $\beta$-function for the strong coupling.
The part $\propto\gs^6\yt^2$ increases it by $\sim 11 \%$ and the part $\propto\gs^4\yt^4$ decreases it by 
$\sim 5 \%$ at this scale compared to the pure QCD contribution $\propto\gs^8$. The non-naive term
gives only a $\sim 0.18 \%$ contribution if we assume the $\gaf$ prescription with a readout point at the external gluon vertices. 
That is $\sim 4\%$ of the total term $\propto\gs^4\yt^4$.
So even if we attached an uncertainty factor of $3$ to the non-naive term the uncertainty is only $\sim 0.6 \%$ 
of the leading term $\propto \gs^8$ at this scale. 
We believe the result presented in this paper to be correct but we nevertheless note here that any deviation due to a different treatment
of $\gaf$ would be phenomenologically irrelevant.
\bk

{\it {\bf Note added 29.08.2016:} In the second version of \cite{Bednyakov:2015ooa} the authors state that there are three possible results for the non-naive part of the four-loop $\beta$-function
\be \left.\f{\beta_{\sss{\als}}^{(4)}}{\als}\right|_{(non-naive)}=\at^2 \as^2 \tr^2 \;  R\; \lb      \f{16}{3}            +  32 \zeta_{3}  \rb\ee
where $R=1,2,3$ depending on the reading point prescription for $\gaf$. $R=1$ corresponds to the external reading point prescription employed in this paper, $R=3$ to the keeping $\gaf$ at there internal
position where they are introduced by the Feynman rules. $R=2$ corresponds to one internal and one external reading point. In \cite{Bednyakov:2015ooa} the self-energies are computed as
massless propagators which allows access to the finite part in addition to the UV divergent part. Only for the internal reading point prescription corresponding to $R=3$ the authors of \cite{Bednyakov:2015ooa}
find a transversal finite part of the self-energy. This suggests that this is the correct result although a formal proof for this treatment of $\gaf$ is still not available.}
          
\section{Conclusions \label{last}}

We have presented an analytical result for the four-loop $\beta$-function of the strong coupling $\gs$ in the gaugeless limit of the SM.
This constitutes an important extension of the well-known QCD result as top-Yukawa coupling is numerically the next important coupling after $\gs$, 
at least at the electroweak scale. Furthermore, this is an important step towards a complete calculation of the four-loop $\beta$-functions of the
gauge couplings in the full SM.

An important feature of this result is the non-naive $\gaf$ contribution $\propto \gs^4 \yt^4$. 
In the pure gauge boson and fermion sector of the SM, given by $\ssL_{\ssst{SU(3)$\times$SU(2)$\times$U(1)}}$, all non-naive contributions cancel in the sum of all diagrams,
making this part of the SM anomaly free. This has been explicitly checked during the calculation of the three-loop $\beta$-functions for the gauge couplings in the SM \cite{PhysRevLett.108.151602,Mihaila:2012pz}.
Here we see that with the inclusion of a scalar field non-naive contributions may appear in higher orders and special care will have to be taken when attempting a complete calculation 
of four-loop $\beta$-functions in the SM.

\section*{Acknowledgements}
I am grateful to K.~G.~Chetyrkin for many useful discussions during the course of this project and
helpful comments on this paper. 

I would like to thank A.~Bednyakov for useful discussions after the first version of this paper
leading to a better understanding of the subtleties of the $\gaf$ treatment in this project.

I also thank  J.~H.~K\"uhn for his support. 

This work has been supported in part by the Deutsche Forschungsgemeinschaft in the
Sonderforschungsbereich/Transregio SFB/TR-9 ``Computational Particle
Physics''.

\bibliographystyle{JHEP}

\bibliography{LiteraturSM}

\providecommand{\href}[2]{#2}\begingroup\raggedright\begin{thebibliography}{10}

\bibitem{Collins:1974da}
J.~C. Collins, {\it {Normal Products in Dimensional Regularization}},  {\em
  Nucl. Phys.} {\bf B92} (1975) 477.

\bibitem{PhysRevLett.108.151602}
L.~N. Mihaila, J.~Salomon, and M.~Steinhauser, {\it {Gauge coupling beta
  functions in the standard model to three loops}},  {\em Phys. Rev. Lett.}
  {\bf 108} (2012) 151602.

\bibitem{Mihaila:2012pz}
L.~N. Mihaila, J.~Salomon, and M.~Steinhauser, {\it {Renormalization constants
  and beta functions for the gauge couplings of the Standard Model to
  three-loop order}},  {\em Phys. Rev. D} {\bf 86} (2012) 096008,
  [\href{http://xxx.lanl.gov/abs/1208.3357}{{\tt arXiv:1208.3357}}].

\bibitem{Bednyakov:2012rb}
A.~Bednyakov, A.~Pikelner, and V.~Velizhanin, {\it {Anomalous dimensions of
  gauge fields and gauge coupling beta-functions in the Standard Model at three
  loops}},  {\em JHEP} {\bf 1301} (2013) 017,
  [\href{http://xxx.lanl.gov/abs/1210.6873}{{\tt arXiv:1210.6873}}].

\bibitem{Chetyrkin:2012rz}
K.~Chetyrkin and M.~Zoller, {\it {Three-loop $\beta$-functions for top-Yukawa
  and the Higgs self-interaction in the Standard Model}},  {\em JHEP} {\bf
  1206} (2012) 033, [\href{http://xxx.lanl.gov/abs/1205.2892}{{\tt
  arXiv:1205.2892}}].

\bibitem{Chetyrkin:2013wya}
K.~Chetyrkin and M.~Zoller, {\it {$\beta$-function for the Higgs
  self-interaction in the Standard Model at three-loop level}},  {\em JHEP}
  {\bf 1304} (2013) 091, [\href{http://xxx.lanl.gov/abs/1303.2890}{{\tt
  arXiv:1303.2890}}].

\bibitem{Bednyakov:2013eba}
A.~Bednyakov, A.~Pikelner, and V.~Velizhanin, {\it {Higgs self-coupling
  beta-function in the Standard Model at three loops}},  {\em Nucl.Phys.} {\bf
  B875} (2013) 552--565, [\href{http://xxx.lanl.gov/abs/1303.4364}{{\tt
  arXiv:1303.4364}}].

\bibitem{Bednyakov:2013cpa}
A.~Bednyakov, A.~Pikelner, and V.~Velizhanin, {\it {Three-loop Higgs
  self-coupling beta-function in the Standard Model with complex Yukawa
  matrices}},  \href{http://xxx.lanl.gov/abs/1310.3806}{{\tt arXiv:1310.3806}}.

\bibitem{Bednyakov:2012en}
A.~Bednyakov, A.~Pikelner, and V.~Velizhanin, {\it {Yukawa coupling
  beta-functions in the Standard Model at three loops}},  {\em Phys.Lett.} {\bf
  B722} (2013) 336--340, [\href{http://xxx.lanl.gov/abs/1212.6829}{{\tt
  arXiv:1212.6829}}].

\bibitem{PhysRevLett.30.1343}
D.~J. Gross and F.~Wilczek, {\it {Ultraviolet Behavior of Non-Abelian Gauge
  Theories}},  {\em Phys. Rev. Lett.} {\bf 30} (1973) 1343--1346.

\bibitem{PhysRevLett.30.1346}
H.~D. Politzer, {\it {Reliable Perturbative Results for Strong Interactions?}},
   {\em Phys. Rev. Lett.} {\bf 30} (1973) 1346--1349.

\bibitem{Jones1974531}
D.~Jones, {\it Two-loop diagrams in yang-mills theory},  {\em Nuclear Physics
  B} {\bf 75} (1974), no.~3 531--538.

\bibitem{Tarasov:1976ef}
O.~Tarasov and A.~Vladimirov, {\it {Two Loop Renormalization of the Yang-Mills
  Theory in an Arbitrary Gauge}},  {\em Sov.J.Nucl.Phys.} {\bf 25} (1977) 585.

\bibitem{PhysRevLett.33.244}
W.~E. Caswell, {\it {Asymptotic Behavior of Non-Abelian Gauge Theories to
  Two-Loop Order}},  {\em Phys.Rev.Lett.} {\bf 33} (1974) 244--246.

\bibitem{Egorian:1978zx}
E.~Egorian and O.~Tarasov, {\it {Two loop renormalization of the QCD in an
  arbitrary gauge}},  {\em Teor.Mat.Fiz.} {\bf 41} (1979) 26--32.

\bibitem{PhysRevD.25.581}
D.~R.~T. Jones, {\it Two-loop $\beta$ function for a ${G}_{1}\times{G}_{2}$
  gauge theory},  {\em Phys. Rev. D} {\bf 25} (1982) 581--582.

\bibitem{Fischler:1981is}
M.~S. Fischler and C.~T. Hill, {\it {Effects of Large Mass Fermions on $M_X$
  and $\sin^2\theta_W$}},  {\em Nucl.Phys.} {\bf B193} (1981) 53.

\bibitem{Jack1985472}
I.~Jack and H.~Osborn, {\it General background field calculations with fermion
  fields},  {\em Nucl. Phys. B} {\bf 249} (1985), no.~3 472--506.

\bibitem{Machacek198383}
M.~E. Machacek and M.~T. Vaughn, {\it Two-loop renormalization group equations
  in a general quantum field theory: (i). wave function renormalization},  {\em
  Nucl. Phys. B} {\bf 222} (1983), no.~1 83--103.

\bibitem{2loopbetayukawa}
M.-x. Luo and Y.~Xiao, {\it {Two loop renormalization group equations in the
  standard model}},  {\em Phys. Rev. Lett.} {\bf 90} (2003) 011601,
  [\href{http://xxx.lanl.gov/abs/hep-ph/0207271}{{\tt hep-ph/0207271}}].

\bibitem{Ford:1992pn}
C.~Ford, I.~Jack, and D.~Jones, {\it {The Standard model effective potential at
  two loops}},  {\em Nucl.Phys.} {\bf B387} (1992) 373--390,
  [\href{http://xxx.lanl.gov/abs/hep-ph/0111190}{{\tt hep-ph/0111190}}].

\bibitem{Fischler1982385}
M.~Fischler and J.~Oliensis, {\it Two-loop corrections to the beta function for
  the higgs-yukawa coupling constant},  {\em Phys. Lett. B} {\bf 119} (1982),
  no.~4 385--386.

\bibitem{Machacek1984221}
M.~E. Machacek and M.~T. Vaughn, {\it Two-loop renormalization group equations
  in a general quantum field theory (ii). yukawa couplings},  {\em Nucl. Phys.
  B} {\bf 236} (1984), no.~1 221--232.

\bibitem{Machacek198570}
M.~E. Machacek and M.~T. Vaughn, {\it Two-loop renormalization group equations
  in a general quantum field theory: (iii). scalar quartic couplings},  {\em
  Nucl. Phys. B} {\bf 249} (1985), no.~1 70--92.

\bibitem{Curtright:1979mg}
T.~Curtright, {\it {Three loop charge renormalization effects due to quartic
  scalar selfinteractions}},  {\em Phys.Rev.} {\bf D21} (1980) 1543.

\bibitem{Jones:1980fx}
D.~Jones, {\it {Comment on the charge renormalization effects of quartic scalar
  selfinteractions}},  {\em Phys.Rev.} {\bf D22} (1980) 3140--3141.

\bibitem{Tarasov:1980au}
O.~Tarasov, A.~Vladimirov, and A.~Y. Zharkov, {\it {The Gell-Mann-Low Function
  of QCD in the Three Loop Approximation}},  {\em Phys.Lett.} {\bf B93} (1980)
  429--432.

\bibitem{Tarasov1980429}
O.~Tarasov, A.~Vladimirov, and A.~Zharkov, {\it The gell-mann-low function of
  qcd in the three-loop approximation},  {\em Physics Letters B} {\bf 93}
  (1980), no.~4 429 -- 432.

\bibitem{3loopbetaqcd}
S.~Larin and J.~Vermaseren, {\it {The Three loop QCD Beta function and
  anomalous dimensions}},  {\em Phys. Lett.} {\bf B303} (1993) 334--336,
  [\href{http://xxx.lanl.gov/abs/hep-ph/9302208}{{\tt hep-ph/9302208}}].

\bibitem{Steinhauser:1998cm}
M.~Steinhauser, {\it {Higgs decay into gluons up to O(alpha**3(s)
  G(F)m**2(t))}},  {\em Phys.Rev.} {\bf D59} (1999) 054005,
  [\href{http://xxx.lanl.gov/abs/hep-ph/9809507}{{\tt hep-ph/9809507}}].

\bibitem{Pickering:2001aq}
A.~Pickering, J.~Gracey, and D.~Jones, {\it {Three loop gauge beta function for
  the most general single gauge coupling theory}},  {\em Phys.Lett.} {\bf B510}
  (2001) 347--354, [\href{http://xxx.lanl.gov/abs/hep-ph/0104247}{{\tt
  hep-ph/0104247}}].

\bibitem{4loopbetaqcd}
T.~van Ritbergen, J.~Vermaseren, and S.~Larin, {\it {The Four loop beta
  function in quantum chromodynamics}},  {\em Phys. Lett.} {\bf B400} (1997)
  379--384, [\href{http://xxx.lanl.gov/abs/hep-ph/9701390}{{\tt
  hep-ph/9701390}}].

\bibitem{Czakon:2004bu}
M.~Czakon, {\it {The Four-loop QCD beta-function and anomalous dimensions}},
  {\em Nucl.Phys.} {\bf B710} (2005) 485--498,
  [\href{http://xxx.lanl.gov/abs/hep-ph/0411261}{{\tt hep-ph/0411261}}].

\bibitem{Cabibbo:1979ay}
N.~Cabibbo, L.~Maiani, G.~Parisi, and R.~Petronzio, {\it {Bounds on the
  Fermions and Higgs Boson Masses in Grand Unified Theories}},  {\em Nucl.
  Phys.} {\bf B158} (1979) 295--305.

\bibitem{Ford:1992mv}
C.~Ford, D.~Jones, P.~Stephenson, and M.~Einhorn, {\it {The Effective potential
  and the renormalization group}},  {\em Nucl. Phys.} {\bf B395} (1993) 17--34,
  [\href{http://xxx.lanl.gov/abs/hep-lat/9210033}{{\tt hep-lat/9210033}}].

\bibitem{Altarelli1994141}
G.~Altarelli and G.~Isidori, {\it Lower limit on the higgs mass in the standard
  model: An update},  {\em Physics Letters B} {\bf 337} (1994), no.~1-2
  141--144.

\bibitem{Bezrukov:2009db}
F.~Bezrukov and M.~Shaposhnikov, {\it {Standard Model Higgs boson mass from
  inflation: two loop analysis}},  {\em JHEP} {\bf 07} (2009) 089,
  [\href{http://xxx.lanl.gov/abs/0904.1537}{{\tt arXiv:0904.1537}}].

\bibitem{Holthausen:2011aa}
M.~Holthausen, K.~S. Lim, and M.~Lindner, {\it {Planck scale Boundary
  Conditions and the Higgs Mass}},  {\em JHEP} {\bf 1202} (2012) 037,
  [\href{http://xxx.lanl.gov/abs/1112.2415}{{\tt arXiv:1112.2415}}].

\bibitem{EliasMiro:2011aa}
J.~Elias-Miro, J.~R. Espinosa, G.~F. Giudice, G.~Isidori, A.~Riotto, et~al.,
  {\it {Higgs mass implications on the stability of the electroweak vacuum}},
  {\em Phys. Lett.} {\bf B709} (2012) 222--228,
  [\href{http://xxx.lanl.gov/abs/1112.3022}{{\tt arXiv:1112.3022}}].

\bibitem{Xing:2011aa}
Z.-z. Xing, H.~Zhang, and S.~Zhou, {\it {Impacts of the Higgs mass on vacuum
  stability, running fermion masses and two-body Higgs decays}},
  \href{http://xxx.lanl.gov/abs/1112.3112}{{\tt arXiv:1112.3112}}.

\bibitem{Bezrukov:2012sa}
F.~Bezrukov, M.~Y. Kalmykov, B.~A. Kniehl, and M.~Shaposhnikov, {\it {Higgs
  Boson Mass and New Physics}},  {\em JHEP} {\bf 1210} (2012) 140,
  [\href{http://xxx.lanl.gov/abs/1205.2893}{{\tt arXiv:1205.2893}}].

\bibitem{Degrassi:2012ry}
G.~Degrassi, S.~Di~Vita, J.~Elias-Miro, J.~R. Espinosa, G.~F. Giudice, et~al.,
  {\it {Higgs mass and vacuum stability in the Standard Model at NNLO}},  {\em
  JHEP} {\bf 1208} (2012) 098, [\href{http://xxx.lanl.gov/abs/1205.6497}{{\tt
  arXiv:1205.6497}}].

\bibitem{Zoller:2012cv}
M.~Zoller, {\it {Vacuum stability in the SM and the three-loop $\beta$-function
  for the Higgs self-interaction}},
  \href{http://xxx.lanl.gov/abs/1209.5609}{{\tt arXiv:1209.5609}}.

\bibitem{Masina:2012tz}
I.~Masina, {\it {Higgs boson and top quark masses as tests of electroweak
  vacuum stability}},  {\em Phys.Rev.} {\bf D87} (2013), no.~5 053001,
  [\href{http://xxx.lanl.gov/abs/1209.0393}{{\tt arXiv:1209.0393}}].

\bibitem{Zoller:2014cka}
M.~F. Zoller, {\it {Standard Model beta-functions to three-loop order and
  vacuum stability}},  \href{http://xxx.lanl.gov/abs/1411.2843}{{\tt
  arXiv:1411.2843}}.

\bibitem{Zoller:2014xoa}
M.~Zoller, {\it {Three-loop beta function for the Higgs self-coupling}},  {\em
  PoS} {\bf LL2014} (2014) 014, [\href{http://xxx.lanl.gov/abs/1407.6608}{{\tt
  arXiv:1407.6608}}].

\bibitem{Zoller:2013mra}
M.~Zoller, {\it {Beta-function for the Higgs self-interaction in the Standard
  Model at three-loop level}},  {\em PoS} {\bf (EPS-HEP 2013)} (2013) 322,
  [\href{http://xxx.lanl.gov/abs/1311.5085}{{\tt arXiv:1311.5085}}].

\bibitem{Buttazzo:2013uya}
D.~Buttazzo, G.~Degrassi, P.~P. Giardino, G.~F. Giudice, F.~Sala, et~al., {\it
  {Investigating the near-criticality of the Higgs boson}},
  \href{http://xxx.lanl.gov/abs/1307.3536}{{\tt arXiv:1307.3536}}.

\bibitem{Bednyakov:2015sca}
A.~V. Bednyakov, B.~A. Kniehl, A.~F. Pikelner, and O.~L. Veretin, {\it {Fate of
  the Universe: Gauge Independence and Advanced Precision}},
  \href{http://xxx.lanl.gov/abs/1507.0883}{{\tt arXiv:1507.0883}}.

\bibitem{Bednyakov:2015ooa}
A.~V. Bednyakov and A.~F. Pikelner, {\it {Four-loop strong coupling
  beta-function in the Standard Model}},
  \href{http://xxx.lanl.gov/abs/1508.0268}{{\tt arXiv:1508.0268}}.

\bibitem{QGRAF}
P.~Nogueira, {\it {Automatic Feynman graph generation}},  {\em J. Comput.
  Phys.} {\bf 105} (1993) 279--289.

\bibitem{Seidensticker:1999bb}
T.~Seidensticker, {\it {Automatic application of successive asymptotic
  expansions of Feynman diagrams}},
  \href{http://xxx.lanl.gov/abs/hep-ph/9905298}{{\tt hep-ph/9905298}}.

\bibitem{Harlander:1997zb}
R.~Harlander, T.~Seidensticker, and M.~Steinhauser, {\it {Complete corrections
  of Order alpha alpha-s to the decay of the Z boson into bottom quarks}},
  {\em Phys.Lett.} {\bf B426} (1998) 125--132,
  [\href{http://xxx.lanl.gov/abs/hep-ph/9712228}{{\tt hep-ph/9712228}}].

\bibitem{Vermaseren:2000nd}
J.~A.~M. Vermaseren, {\it {New features of FORM}},
  \href{http://xxx.lanl.gov/abs/math-ph/0010025}{{\tt math-ph/0010025}}.

\bibitem{Tentyukov:2007mu}
M.~Tentyukov and J.~A.~M. Vermaseren, {\it {The multithreaded version of
  FORM}},  \href{http://xxx.lanl.gov/abs/hep-ph/0702279}{{\tt hep-ph/0702279}}.

\bibitem{COLOR}
T.~Van~Ritbergen, A.~Schellekens, and J.~Vermaseren, {\it Group theory factors
  for feynman diagrams},  {\em International Journal of Modern Physics A} {\bf
  14} (1999), no.~1 41--96.

\bibitem{MATAD}
M.~Steinhauser, {\it {MATAD: A program package for the computation of massive
  tadpoles}},  {\em Comput. Phys. Commun.} {\bf 134} (2001) 335--364,
  [\href{http://xxx.lanl.gov/abs/hep-ph/0009029}{{\tt hep-ph/0009029}}].

\bibitem{Vermaseren:1994je}
J.~A.~M. Vermaseren, {\it Axodraw},  {\em Comput. Phys. Commun.} {\bf 83}
  (1994) 45--58.

\bibitem{Smirnov:2008iw}
A.~Smirnov, {\it {Algorithm FIRE -- Feynman Integral REduction}},  {\em JHEP}
  {\bf 0810} (2008) 107, [\href{http://xxx.lanl.gov/abs/0807.3243}{{\tt
  arXiv:0807.3243}}].

\bibitem{Smirnov:2014hma}
A.~V. Smirnov, {\it {FIRE5: a C++ implementation of Feynman Integral
  REduction}},  {\em Comput.Phys.Commun.} {\bf 189} (2014) 182--191,
  [\href{http://xxx.lanl.gov/abs/1408.2372}{{\tt arXiv:1408.2372}}].

\bibitem{Smirnov:2013eza}
A.~V. Smirnov, {\it {FIESTA 3: cluster-parallelizable multiloop numerical
  calculations in physical regions}},  {\em Comput.Phys.Commun.} {\bf 185}
  (2014) 2090--2100, [\href{http://xxx.lanl.gov/abs/1312.3186}{{\tt
  arXiv:1312.3186}}].

\bibitem{Misiak:1994zw}
M.~Misiak and M.~M{\"u}nz, {\it {Two loop mixing of dimension five flavor
  changing operators}},  {\em Phys. Lett.} {\bf B344} (1995) 308--318,
  [\href{http://xxx.lanl.gov/abs/hep-ph/9409454}{{\tt hep-ph/9409454}}].

\bibitem{beta_den_comp}
K.~G. Chetyrkin, M.~Misiak, and M.~M{\"u}nz, {\it {Beta functions and anomalous
  dimensions up to three loops}},  {\em Nucl. Phys.} {\bf B518} (1998)
  473--494, [\href{http://xxx.lanl.gov/abs/hep-ph/9711266}{{\tt
  hep-ph/9711266}}].

\bibitem{Larin:1993tq}
S.~Larin, {\it {The Renormalization of the axial anomaly in dimensional
  regularization}},  {\em Phys.Lett.} {\bf B303} (1993) 113--118,
  [\href{http://xxx.lanl.gov/abs/hep-ph/9302240}{{\tt hep-ph/9302240}}].

\bibitem{Korner:1991sx}
J.~G. Korner, D.~Kreimer, and K.~Schilcher, {\it {A Practicable gamma(5) scheme
  in dimensional regularization}},  {\em Z. Phys.} {\bf C54} (1992) 503--512.

\bibitem{ATLAS:2014wva}
ATLAS, CDF, CMS, and D0, {\it {First combination of Tevatron and LHC
  measurements of the top-quark mass}},
  \href{http://xxx.lanl.gov/abs/1403.4427}{{\tt arXiv:1403.4427}}.

\bibitem{Aad:2015zhl}
{\bf ATLAS, CMS} Collaboration, G.~Aad et~al., {\it {Combined Measurement of
  the Higgs Boson Mass in $pp$ Collisions at $\sqrt{s}=7$ and 8 TeV with the
  ATLAS and CMS Experiments}},  {\em Phys. Rev. Lett.} {\bf 114} (2015) 191803,
  [\href{http://xxx.lanl.gov/abs/1503.0758}{{\tt arXiv:1503.0758}}].

\bibitem{Bethke:2012jm}
S.~Bethke, {\it {World Summary of $\alpha_s$ (2012)}},  {\em
  Nucl.Phys.Proc.Suppl.} {\bf 234} (2013) 229--234,
  [\href{http://xxx.lanl.gov/abs/1210.0325}{{\tt arXiv:1210.0325}}].

\end{thebibliography}\endgroup

\end{document}